\documentclass{aa}
\usepackage{natbib}
\usepackage{graphicx}

\newcommand{\lsim}{\raisebox{-0.3ex}{\mbox{$\stackrel{<}{_\sim} \,$}}}

\def\be{\begin{equation}}
\def\ee{\end{equation}}

\def\gtrsim{\raisebox{-0.3ex}{\mbox{$\stackrel{>}{_\sim} \,$}}}

\begin{document}

\title{Vacuum gap model for PSR B0943$+$10}

\author{Janusz A. Gil
        \inst{1}
        \and
        George I. Melikidze
        \inst{1,2}
        \and
        Dipanjan Mitra
        \inst{3}}

\institute{Institute of Astronomy, University of Zielona G\'ora,
Lubuska 2, 65-265, Zielona G\'ora, Poland \and Center for Plasma
Astrophysics, Abastumani Astrophysical Observatory, Al.Kazbegi
ave. 2a, Tbilisi 380060, Georgia \and Max-Planck Institute f\"ur
Radioastronomie, Auf dem H\"ugel 69, D-53121, Bonn, Germany}


\authorrunning{Gil, Melikidze, Mitra.}

\titlerunning{Vacuum gap model for PSR B0943$+$10}

\date{Received / Accepted }

\abstract{PSR B0943$+$10 is known to show remarkably stable
drifting subpulses, which can be interpreted in terms
of a circumferential motion of 20 sparks, each completing one
circulation around the periphery of the polar cap in 37 pulsar
periods. We use this observational constraint and argue that the
vacuum gap model can adequately describe the observed drift
patterns. Further we demonstrate that {\em only} the presence of
strong non-dipolar surface magnetic field can favor such vacuum
gap formation. Subsequently, for the first time we are able to
constrain the parameters of the surface magnetic field, and model
the expected magnetic structure on the polar cap of PSR B0943$+$10
considering the inverse Compton scattering photon dominated vacuum
gap. } \maketitle

\section{Introduction}

The fundamental problem of pulsar research is lack of consensus
about the radio emission mechanism, except that the
electron-positron pair production is an essential ingredient,
to explain the observed coherent radiation. Moreover,
the observed phenomenon of drifting subpulses serves as an
excellent diagnostic tool to investigate the mechanism of pulsar
radio emission. PSR B0943+10 with period $P=1.09$~s and period
derivative $\dot{P}=3.5\cdot 10^{-15}$ s/s exhibits long and
stable sequences of drifting subpulses (Deshpande \& Rankin 1999,
2001, hereafter DR99 and DR01, Asgekar \& Deshpande 2001). This
pulsar is observable only at frequencies below 800 MHz (high
frequency radiation most probably misses the line-of-sight due to
specific observing geometry) and drifting subpulses with similar
features are observed at 35, 111 and 430 MHz, implying broad-band
nature of the phenomenon. This suggests that drifting subpulses in
this pulsar are associated with a stable system of emission
subbeams, rotating around the pulsar magnetic axis. The aliasing
resolved analysis of the harmonic-resolved fluctuation spectra
related to drifting subpulses yielded in precise estimates of
periodicities of phase and amplitude modulation, namely
$P_3=1.87~P$ and $\hat{P}_3=37~P$, respectively (DR01). These two
periodicities are harmonically related and are used to determine
the number of subpulse beams $N=\hat{P}_3/P_3=20$ contributing to
the observed modulations. The subpulse ``drift-bands'' clearly
visible in the observed drift pattern (Backer 1973) are separated
vertically by $\hat{P}_3$ (in pulsar periods) and horizontally by
$P_2$ (in pulse longitude). The latter is estimated as being about
$11^{\circ}$ and can be converted into the magnetic azimuth angle
$\eta$ separating adjacent subbeams. This requires knowledge of
the inclination angle $\alpha$ and the impact angle $\beta$, which
for this pulsar is estimated from the polarization measurements
and from the viewing geometry that leads to missing radiation
above $\sim 800$~MHz. Using $\alpha=11^{\circ}.64$ and
$\beta=-4.31^{\circ}$ (DR01), one gets $\eta=18^\circ$, which in
turn gives $N=360^\circ/\eta=20$ subbeams rotating at the beam
periphery of PSR B0943$+$10.

 Karastergiou et al.(2001) observed simultaneously subpulses in single
pulses of PSR B0329+34 at 1.4 and 2.7 GHz. Careful analysis of these
observations suggests strongly that a single plasma column fed 
by a single entity at the bottom of the pulsar magnetosphere
is responsible for the subpulse observed
across the pulsar spectrum, which can be interpreted naturally in terms
of sparks operating on the polar cap surface.
This picture is probably more general and can be
applied to other pulsars, including those with drifting subpulses.
\citet{dr99} noticed that the subpulse drift phenomenon observed
in PSR B0943$+$10 is consistent with the model proposed by
Ruderman \& Sutherland (1975, hereafter RS75), in which the
subbeams are associated with sparking discharges operating within
the vacuum gap (VG) developed just above the polar cap. The
observed subpulse drift rate in this model arises from the speed
of circulation of sparks around the magnetic axis, which is
determined by rate of the ${\bf E}\times{\bf B}$ drift of 
electron-positron plasma confined to the isolated spark filament within 
the VG. If the planes of surface magnetic field lines are more or
less perpendicular to the polar cap boundary, then the peripherial
sparks will slowly rotate around the magnetic axis. \citet{dr99}
demonstrated clearly that the non-corotating subbeams associated with
drifting subpulses actually lag the stellar rotation. This finding
strenghtens the VG plasma ${\bf E}\times{\bf B}$ drift hypothesis.  
However, the formation of 
VG itself suffered criticism owing to the problem of
binding energy of ions, not being high enough to be retained on
the stellar surface (Hillenbrandt \& M\"{u}ller 1976, K\"{o}ssl et
al 1988). Morover, the association of drifting subpulses with sparks
was questioned based on the fact that sparks will move
towards the magnetic pole at a speed much faster than the
${\bf E}\times{\bf B}$ drift velocity, thus quenching the stable
drifting patterns \citep{cr77,fr82}. In this paper we demonstrate that
indeed VG can form in PSR B0943$+$10 if the surface magnetic field
is very strong and non-dipolar in nature. For the first time we
are able to constrain the parameters of the surface magnetic field
assuming that the VG is dominated by the inverse Compton
scattering photons. We also demonstrate that the embarrassing
problem with rapid spark motion towards the pole will not occur
due to specific geometry of the surface magnetic field in the VG,
at least in PSR B0943$+$10.

\section{Constraining surface field in PSR B0943$+$10}

\subsection{Vacuum Gap formation}

Gil \& Mitra (2001, hereafter GM01) argued that VG can form in
neutron stars with very strong non-dipolar surface magnetic field
$B_s=b\cdot B_d$,  where $B_d=2\cdot
10^{12}(P\cdot\dot{P}_{-15})^{0.5}$~G is the inferred dipolar
field $(B_d=4\times 10^{12}$~G in this case) and
$\dot{P}_{-15}=\dot{P}/10^{-15}$. For surface magnetic fields with
$b\gg 1$, the radius of curvature of field lines ${\cal R}$ should
be relatively small, with normalized values ${\cal R}_6={\cal
R}/R<< 100$, where $R=10^6$~cm is the neutron star radius. GM01
considered strong magnetic fields $B>0.1~B_q$, where
$B_q=4.4\times 10^{13}$~G is termed as the quantum critical field.
In such strong magnetic fields, high energy photons with frequency
$\omega$ produce electron-positron pairs {\it near the kinematic
threshold} with photon energy $\hbar\omega=2mc^2/\sin\theta$,
where $\sin\theta=l_{ph}/{\cal R}$, $l_{ph}$ is the mean photon
path for pair formation. The typical photon energy is
$\hbar\omega=1.5\cdot\hbar\gamma^3c/{\cal R}$ in the case of
curvature radiation (CR) seed photons (e.g. RS75) and
$\hbar\omega=2\gamma\hbar eB/mc$ in the case of resonant inverse
Compton scattering (ICS) seed photons \citep{zql97}; see GM01 for
details of the Near Threshold Vacuum Gap (NTVG) formation.

The condition for VG formation in neutron stars with
${\bf\Omega}\cdot{\bf B}<0$ can be written in the form
$T_i/T_s>1$, where $ T_i=6\times
10^5b^{0.73}(P\cdot\dot{P}_{-15})^{0.36}~K $ is the critical
temperature above which $^{56}$Fe ions cannot be bound
\citep{as91,um95}. The surface temperature $T_s=\left(k\cdot
e\Delta V\dot{N}/\sigma\pi r_p^2\right)^{1/4}$, where $
r_p=b^{-0.5}10^4P^{-0.5}~{\rm cm} $ is the actual polar cap
radius, $ \dot{N}_{\rm GJ}=\pi r_p^2B_s/(eP) $ is the kinematic 
Goldreich-Julian flux
through the polar cap surface, $ \Delta V=(2\pi/cP)B_s h^2 $ is
the potential drop across the polar gap. The {\it efficiency
coefficient} $k$ in $\Delta V$ describes departure from black body
conditions on the polar cap surface; $k=q_{rad}/q_{heat}<1$, where
$q_{rad}=\sigma T_s^4$, $q_{heat}=-({\bf\Omega}\cdot{\bf B_s}/2\pi
c)\cdot\Delta V$ and $\sigma=5.67\times 10^{-5}{\rm erg\
cm}^{-2}K^{-1}s^{-1}$. One can show that at the neutron star
surface below the sparking gap the value of $k$ can be as low as
0.3 (Gil 2002, in preparation). Considering the CR seed photons as
a source of electron-positron pairs one obtains the height of the
CR dominated NTVG in the form $ h_{CR}=(3\times 10^3){\cal
R}_6^{0.29}b^{-0.43}P^{0.21}\dot{P}_{-15}^{-0.21}~{\rm cm} $
(GM01; their eq.~6)\footnote{ Note the typographical error in GM01
eq.~6 for powers of $P$ and $\dot{P}$ which is corrected here.}.
Using it for VG condition $T_{i}/T_{s}
> 1$, we get the relation between $b$ and ${\cal R}_6$ in the form
\be b= 15.6 \times {\cal R}_6^{0.31}~~~({\rm for}~k=0.5),
\label{eq1} \ee and $b=20\times{\cal R}^{0.31}_6$ for $k=1.0$. In
the case of the resonant-ICS dominated gap, the height of the NTVG
is $ h_{ICS}=(5\times 10^3){\cal
R}_6^{0.57}b^{-1}P^{-0.36}\dot{P}_{-15}^{-0.5}~ {\rm cm}$ (GM01;
their eq.~12) and the corresponding relation for gap formation is,

\begin{eqnarray}
b &= 5.1 \times {\cal R}_6^{0.38}~~~({\rm for}~k=0.5),\\
b &= 6.4 \times {\cal R}_6^{0.38}~~~({\rm for}~k=1.0).\label{eq2}
\end{eqnarray}

We argue in section 4 that from our modelling CR-NTVG is not
likely to form in PSR B0943$+$10 and that for ICS-NTVG the value
of the coefficient $k$ is slightly lower than unity. Hereafter we
only consider ICS-NTVG for further calculations.

\subsection{Complexity parameter}

Following the original idea of RS75, Gil \& Sendyk (2000;
hereafter GS00) argued that the polar cap is populated with
$N_{max}\sim a^2$ sparks with characteristic dimension ${\cal
D}\sim h$, separated from one another by a distance $d \sim h$, where
$a=r_p/h$ is the {\it complexity parameter} (see also Fan et al.
2001). Keeping in mind the grazing viewing trajectory in PSR
B0943$+$10, we consider an arrangement of $N=20<N_{max}$ sparks on
a periphery of the polar cap in this pulsar. Since the exact
separation $d\sim h$ of adjacent sparks is unknown, one can write
$ d=h(1+\delta)$ where $|\delta|\ll 1$. 
 We will arbitrarily choose the value of $|\delta|=0.25$, which
determines the parameter space marked by the grey area in Fig. 1
(note that if $|\delta|=0$ then the parameter space reduces to single
line midway between lines (4) and (5), which is both unrealistic
and inconvenient for further analysis).
We can thus write
$N\cdot{\cal D}+N\cdot d=20h(2+\delta)=2\pi(r_p-h)$ or $
a=7.34+3.18\ \delta$. Using the ICS dominated gap height
$h=h_{ICS}$ we obtain the relation between $b$ and ${\cal R}_6$ in
the form $b^{0.5}{\cal R}_6^{-0.57}=2+0.86\ \delta$, which gives
two limiting relations of the form
\begin{eqnarray}
b &= 4.9 \times {\cal R}_6^{1.14}&~~~({\rm for}~\delta=+ 0.25) , \\
b &= 3.2 \times {\cal R}_6^{1.14}&~~~({\rm for}~\delta=- 0.25) ,
\label{eq4}
\end{eqnarray}
corresponding to lines (4) and (5) in Fig.~1, respectively.
Any
point $(b, R_6)$ between these lines corresponds to the VG with 20
sparks of diameter ${\cal D}=h$, separated from each other by
$d=h(1+\delta)$ (where $-0.25<\delta<+0.25$),
occupying a peripheral ring-like area on the
polar cap of PSR B0943+10. Although the maximum possible number of
sparks on the polar cap of this pulsar is $N_{max}\sim a^2\sim 50$
(GS00, Fan et al. 2001), only the 20 outermost sparks can contribute to
the observed radiation due to the grazing line-of-sight geometry. 
Moreover, lack of conditions for vacuum
gap formation towards the central parts of the polar cap can
result in the hollow-cone structure of emission beams (see next
section; Fig.~3).

\subsection{The ${\bf E}\times{\bf B}$ drift rate}

The isolated columns of spark plasma within VG perform a slow
${\bf E}\times{\bf B}$ drift in the direction perpendicular to the
planes of magnetic field lines, which in turn should be
perpendicular to the polar cap edge.  Following RS75, one can
attempt to estimate the speed of circumpferential motion of each
spark $v_{d}=cE_{\perp}/B_{s}$, where $E_{\perp}\sim 2\Delta V/r_p$
(to within an uncertainty factor of less than 4) and $\Delta V=(2\pi/cP)B_sh^2$
is the potential drop within the gap.
Thus, the speed of ${\bf E}\times{\bf B}$ circulation $v_d=(4\pi /P)(h^2/r_p)$.
 In the case of
PSR B0943+10, $h \sim 500$ cm (see Fig.3), $r_p \sim 3300$ cm and
$v_d \sim 850$ cm/s. The time interval after which each spark
completes full ($360^{\circ}$) circulation around the polar cap perimeter
is $\hat{P}_3 \sim 2\pi r_p/v_{d}$=0.5$P(r_p/h)^2$, which is about 22$P$
in this case. Given the uncertainty in $E_{\perp}$, this
is consistent with to the observed value $\hat P_3=37P$ ( which implies
$v_d=2\pi r_p/\hat{P}_3\sim 510$ cm/s, close to $810$ cm/s obtained above
given the uncertainty in the theoretical determination of $E_{\perp}$).

\begin{figure}
{\includegraphics[height=8cm, width=5.5cm, angle=-90]{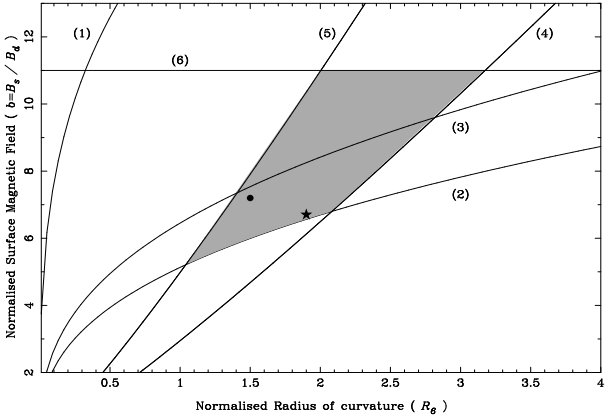}}
\caption[]{Plot of the normalized surface magnetic field
$b=B_s/B_d$ versus normalized radius of curvature ${\cal
R}_6={\cal R}/R$. Lines (1) to (6) in the above figure correspond
to eqs. (1) to (6) in the text. The allowed parameter space for PSR
B0943$+$10 is represented by the gray area. The two points with
co-ordinates $({\cal R}_6,b)$ marked by $\star$ and $\bullet$
represent two modelled cases (1.5, 7.2) and (1.9, 6.7),
respectively, described in section 3 and presented in Figs.~2 and
3.} \label{fig1}
\end{figure}

\subsection{Parameter space}

The surface magnetic field $B_s\sim B_q=4.4\times 10^{13}$~G
represents the so-called photon splitting level, above which the
electron-positron plasma cannot be produced (see Zhang \& Harding
2000 for review). Hence, yet another constraint can be imposed for
PSR B0943+10 \be b \leq \frac{B_q}{B_d} \sim 11 . \label{eq5} \ee
Thus, the parameter space for possible values of $b$ and ${\cal
R}_6$ is marked by the gray area enclosed by lines (6), (5), (4)
and (2)\footnote{The line (2) in Fig.~1 corresponds to the case
when the actual value of the efficiency parameter
$k=q_{rad}/q_{heat}=0.5$ is adopted. For $k=1$ the line (2) should
be replaced by the line (3). Therefore, the region between lines
(2) and (3) corresponds to $0.5<k<1$, while the region above line
(3) corresponds to $k=1$ (for more detailed discussion of the
parameter $k$ see Gil (2002), in preparation).} in Fig.~1. Within
this area the ICS-dominated NTVG can exist, with 20 {\bf equidistant}
sparks circulating around the perimeter of the polar cap in the
time interval $\hat{P}_3$ close to 37 pulsar periods. The CR-NTVG
represented by line (1) correspond to the case $k=0.5$ (for
$k=1.0$ the limiting line $b=20\times{\cal R}_6^{0.31}$ almost
coincides with the ordinate axis in Fig.~1; see also section~4).
While we are not able to exclude the possibility of CR-NTVG
formation, we find modelling of implied magnetic structure
extremely difficult as we describe in section 3 and 4. As seen
from Fig.~1, the surface magnetic field $B_s$ exceeds $2\times
10^{13}$~G $(b>3)$ and the radius of curvature ${\cal R}$ of
surface field lines is about $(1-3)\times 10^6$~cm $(1\lsim{\cal
R}_6\lsim 3)$. The two filled symbols ($\star$ and $\bullet$)
represent two model examples of the surface magnetic field
structure which are discussed in the next section.

 \section{Modelling the surface magnetic field}

We have argued that the observations of regularly drifting
subpulses in PSR B0943+10 imply that ICS-dominated NTVG can form
in this pulsar with a strong non-dipolar surface magnetic field.
Further from Fig~1, we notice that the two parameters lie in the
range $3\lsim b\lsim 11$ and $0.33\lsim\rho_c\lsim 1$, where we
introduce the normalized curvature $\rho_c=1/{\cal R}_6$.
We now examine by means of numerical modelling whether
such a structure of the surface magnetic field is conceivable. We
assume that the actual surface magnetic field is a superposition
of the star centered global dipole moment ${\bf d}$ and the crust
anchored dipole moment ${\bf m}$ placed at ${\bf R}_m$, whose
influence results in small scale deviations of the surface field from
the pure global dipole. The resultant surface magnetic field ${\bf
B}_s={\bf B}_d+{\bf B}_m$, where $|{\bf B}_d|\sim 2d/r^3$ and
$|{\bf B}_m|\sim 2m/|{\bf r}-{\bf R}_m|^3$, where ${\bf r}$ is the
radius vector. For clarity of graphic presentation we use
the normalization $R_0=|{\bf R}_m|/R$ and the strength of $m=|{\bf
m}|$ is expressed in units of $d=|{\bf d}|$.

The detailed numerical formalism of modelling the actual surface
magnetic field within the above model was developed by
\citet{gmm02} and we now apply it to PSR B0943+10. This
accompanying paper gives details of  modelling of the surface
magnetic field in pulsars (a review concerning the crust origin
surface magnetic field in pulsars is also included in this paper).
To obtain the equations of resultant open magnetic field lines
(along which a high unipolar potential drop can develop) we solve
the system of differential equations $d\theta/dr=B_\theta/(rB_r)$
and $d\phi/dr=B_\theta(rB_r\sin\theta)$, with the initial
condition $B_m(r=5R)=0$ (since $B_m/B_d\sim (m/d)(r/|{\bf r}-{\bf
R}_m|)^3$, then for $m/d\sim 10^{-4}$ the ratio $B_m/B_d\sim
10^{-4}$ for $r\gtrsim 5R$). Thus, we define a bundle of the open
dipolar field lines at the altitude $r=5R$ and then trace the
lines of resultant magnetic field ${\bf B}_s$ down to the stellar
surface $(r=R)$. Consequently, we obtain the boundary of the
physical polar cap as well as the actual surface magnetic field
above the polar cap. It follows from the symmetry suggested by the
observed patterns of drifting subpulses and their spectral
analysis (DR99, DR01) that the local dipole ${\bf m}$ should lie
in the fiducial plane containing both the pulsar spin axis and the
magnetic axis. Without loss of generality, we placed for
simplicity the local dipole axis at the polar cap center. Since
$B_s\gg B_d$ in PSR B0943$+$10 then in order to increase $B_d$ to
values exceeding $10^{13}$~G the magnetic moments ${\bf d}$ and
${\bf m}$ must have the same polarity.

\begin{figure}
 {\includegraphics[height=9.5cm, width=8cm]{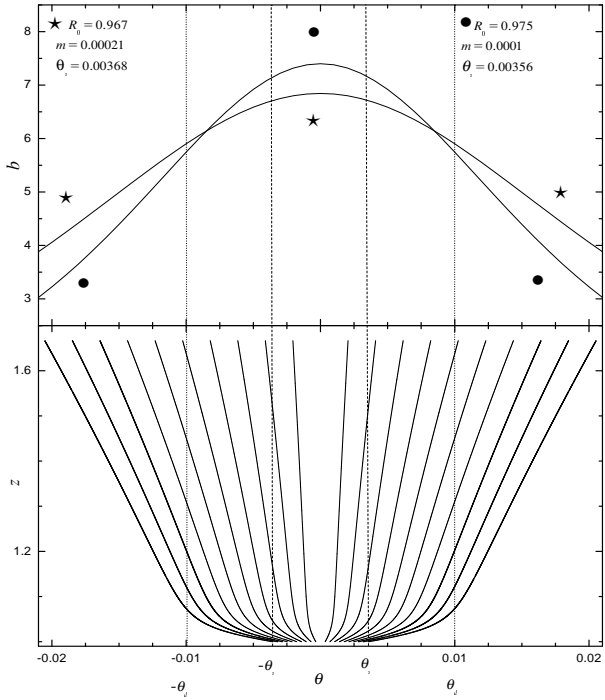}}
\caption[]{Structure of the surface magnetic field ${\bf B}_s$ for
a superposition of the global star centered dipolar moment and
crust anchored dipole moment described by: $m=2.85\times
10^{-4}~d$ and $R_0=0.967~R$ ($\star$) and $m=1.73\times
10^{-4}~d$ and $R_0=0.975~R$ ($\bullet$), respectively. The
horizontal axis is labelled by the azimuthal angle $\theta$
(magnetic co-latitude). The vertical axis is labelled by the
normalized surface magnetic field $b=B_s/B_d$ in the upper panel,
and by the normalized altitude $z=r/R$ in the lower panel. The
angular size of the canonical ($\theta_d$--dipolar) and the actual
($\theta_s$--non-dipolar) polar cap is marked by dotted and dashed
vertical lines, respectively. The converging magnetic field lines
presented in the lower panel correspond to both cases, but, given
the proximity of the two structures, cannot be distinguished
within the graphic resolution.} \label{fig2}
\end{figure}

Fig.~2 presents results of calculations for two cases:
$m=2.1\times 10^{-4}d$ and $R_0=0.967R$ (marked as $\star$), and
$m=1.0\times 10^{-4}d$ and $R_0=0.975R$ (marked as $\bullet$). The lower
panel presents the actual surface magnetic fields (as a function
of normalized altitude $z=r/R$) for both cases, however, the
proximity of the structures makes it difficult to distinguish them
in the plotted scale. The abscissa is labelled by the azimuthal
angle $\theta$ (magnetic colatitude), which measures the polar cap
radius in radians. For a purely dipolar magnetic field
$r_p=R\cdot\sin\theta_d\approx 10^4P^{-1/2}$~cm, which corresponds
to $|\theta_d|\sim 0.01$ radians (marked by two vertical dotted
lines). The actual polar cap is much narrower, with
$|\theta_s|\sim 0.004$ radians (marked by two vertical dashed
lines) or $r_s\sim 0.4r_d$. Near the last open magnetic field
lines (thick lines in the lower panel) $b=6.7$ ($\bullet$) and
$b=7.2$ ($\star$), respectively.

Fig.~3 shows the average normalized curvature $\rho_c=1/{\cal
R}_c$ of the two outermost (thick) magnetic field lines shown in
Fig.~2. The horizontal axis is labelled by $\zeta=z-1$, where
$z=r/R$ (Fig.~2). The dashed vertical line corresponds to the
characteristic gap height $h\approx 500$~cm. As follows from the
parameter space marked by the gray area in Fig.~1, the normalized
radius of curvature $1\lsim{\cal R}_6\lsim 3$ and thus the
normalized curvature $0.33 \lsim \rho_c \lsim 1$. Therefore,
conditions for the formation of the ICS-NTVG are satisfied within
the gray area marked in Fig.~3. This corresponds to an annular
region at the periphery of the polar cap (with size comparable
with the gap height $h_{ICS}\sim 0.15 r_p)$, and can accommodate
only one ring of 20 drifting sparks (with
${\cal D}\sim h_{ICS}$). Within the central region inside this
ring, the polar gap with sparking discharges cannot develop. Thus,
either the central part of the beam of PSR B0943$+$10 is
radio-quiet (hollow-cone) or its radio emission is driven by the
stationary free-flow (e.g. Sharleman et al. 1978) instead of
nonstationary spark-associated flow (RS75).

As it can be seen from Fig.~3, the crust-origin magnetic field
decays rapidly with altitude and at $\zeta>0.3$ ($z>1.3$) the
normalized curvature $\rho_c\simeq 0.007$ (${\cal R}_6\simeq
150$), suggesting almost for purely dipolar field.
Moreover, the sign of $\rho_c$ in the region of dipolar field
lines is negative, while within the gap (and its vicinity)
$\rho_c$ is positive. Thus, the geometry of the surface magnetic
fields lines within the gap is converging, implying that the
radius of curvature is directed towards the pole (opposite to the
case of diverging dipolar field lines). Therefore, the magnetic
field driven spark discharges will tend to occupy the peripheral
areas of the polar cap, instead of moving rapidly towards the pole
\citep{cr77,fr82}. This should have a stabilizing effect on the
arrangement of 20 sparks circulating around the perimeter of the
polar cap in PSR B0943$+$10.

\begin{figure}
 {\includegraphics[height=7.5cm, width=8cm]{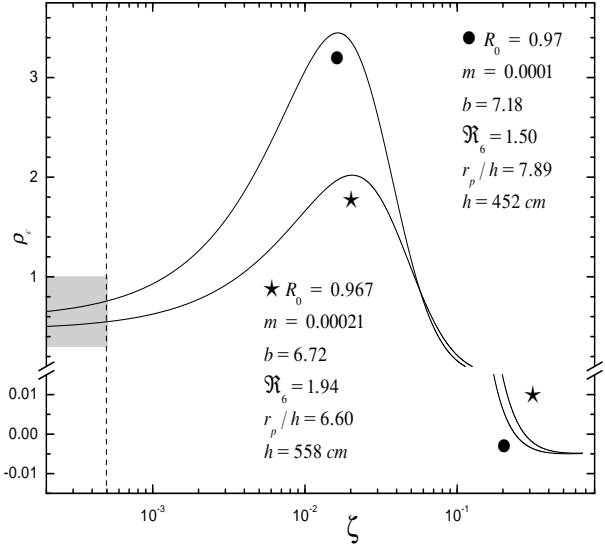}}
\caption[]{Plot of the normalized curvature $\rho_c$ averaged for
the outermost open magnetic field lines (thick lines in Fig.~2)
versus the normalized altitude expressed in the form $\zeta=z-1$,
where $z=r/R$ (Figs.~2 and 3). The parameters of the modelled
points marked by $\star$ and $\bullet$ are given. The gray area
indicates values of $\rho_c$ for which the ICS-NTVG can form in
PSR B0943$+$10. The vertical dashed line represents the VG height
$h\sim 500$~cm.} \label{fig3}
\end{figure}

\section{Summary}

In this paper we explored consequences of the vacuum gap model
interpretation of drifting subpulses in PSR B0943$+$10, where 20
sparks move circumferentially around the perimeter of the polar
cap, each completing one circulation in 37 pulsar periods (DR99,
DR01). Both the characteristic spark dimension $\mathcal{D}$ and
mutual distance $d$ between adjacent sparks are assumed to be
about the polar gap height $h$. The formation of the vacuum gap
requires extremely strong non-dipolar surface magnetic (GM01). We
considered both the CR and resonant ICS seed photons as sources of
electron-positron pairs and determined the parameter space for the
surface magnetic field structure in each case. For the CR-NTVG
($k=0.5$) the surface magnetic field strength $B_{s}>2\times
10^{13}$~G and the radius of curvature of surface field lines
$6\times 10^{4}\mathrm{cm}<\mathcal{R}<12\times
10^{4}\mathrm{cm}$, while for the resonant ICS-NTVG $B_{s}>2\times 10^{13}$%
~G and $10^{6}\mathrm{cm}<\mathcal{R}<3\times 10^{6}\mathrm{cm}$
(in both cases $B_{s}<B_{q}\sim 4.4\times 10^{13}$~G). Our
modelling favours ICS-NTVG formation (refer  Fig.~1) although
CR-NTVG cannot be excluded. The ICS photon driven vacuum gap
supported by the magnetic field
structure determined by this parameter space guarantees a system of 20 sparks with $%
\mathcal{D}\sim d\sim h_{ICS}$, circulating around the perimeter
of the polar cap by means of the $\mathbf{E}\times \mathbf{B}$
drift in about 37 pulsar periods. This phenomenon is most probably
observed in the form of drifting subpulses in PSR B0943$+$10.

We modelled the magnetic field structure using a simple and quite
general model in which the actual surface magnetic field at the
polar cap is a superposition of global core-anchored dipole
$\mathbf{B}_{d}$ (inferred from pulsar spin down) and local
crust-anchored dipole $\mathbf{B}_{m}$ \citep{gmm02}. Since
$B_{d}=6.4\times 10^{12}(P\cdot \dot{P})^{1/2}\mathrm{G}=4\times
10^{12}$~G, then in order to obtain $\mathbf{B}_{s}\sim (2\div
3)\times 10^{13}$~G one needs to assume $B_{m}\gg B_{d}$ and the
same polarity of both components. Following the symmetry suggested
by observed patterns of drifting subpulses in PSR B0943$+$10, we
placed the
local dipole axis at the polar cap center. Two examples of model solutions $%
(b,\mathcal{R})$ are shown in Fig.~1 marked as $\star$ and $\bullet$
within the gray area. More details concerning these two solutions
are presented in Figs.~2 and 3. Our modelling yields a number of
important conclusions: (i) The conditions for the formation of the
ICS-NTVG are satisfied only at peripheral ring-like region of the
polar cap which can just accommodate a system of 20
$\mathbf{E}\times \mathbf{B}$ drifting sparks with
$\mathcal{D}\sim d\sim h_{ICS}$. (ii) The surface magnetic field
lines within the actual gap are converging, which stabilizes the
$\mathbf{E}\times \mathbf{B}$ drifting sparks by preventing them
from rushing towards the pole (as in the case of diverging dipolar
field). (iii) No model solutions could be obtained close to or
above the line (3) in Fig.~1 (especially for
$\mathcal{R}_{6}>1.5$), corresponding to the efficiency
parameter$^2$ $k=1$. This means that $k\lsim 0.8$ ($\bullet$ in
Fig.~1), at least in PSR B0943$+$10. (iv) We find no model
solutions in the range $b\sim$ 7 to 10 and $\mathcal{R}_{6}\sim$ 0.06
to 0.12, corresponding to the CR-NTVG parameter space (not shown in
Fig.~1). This favors the ICS-NTVG in PSR B0943$+$10. We emphasize
that conclusions (ii)-(iv) result from a specific dependence of
$b$ on ${\cal R}_b$ following from central position of the local
surface dipole ${\bf B}_m$, which seems to be justified in this
case by apparent symmetries inferred from the observational data
of PSR B0943$+$10.

 The sparks inherent to the vacuum gap model not only
provide a natural explanation for drifting subpulses but they also
imply a mechanism of their radio emission. Melikidze et al. (2000)
proposed a spark-associated model for coherent pulsar radio
emission by means of curvature radiation of charged relativistic
solitons and we will apply this model to the ICS-NTVG in PSR
B0943$+$10 in the forthcoming paper.

 \begin{acknowledgements}
This paper is supported in part by the Grant 2~P03D~008~19 of the
Polish State Committee for Scientific Research. We are grateful to
B. Zhang and B. Rudak for helpful comments. 
DM would like to thank Institute of Astronomy, University of Zielona
G\'ora, for support and hospitality during his visit to the
institute, where this work was started.
We thank E. Gil for technical assistance and  Floris van der Tak
for careful reading of the manuscript.
 \end{acknowledgements}


\begin{thebibliography}{}

\bibitem[Abrahams \& Shapiro(1991)]{as91} Abrahams A.M., Shapiro S.L. 1991, ApJ, 374, 652
\bibitem[Asgekar \& Deshpande(2001)]{ad01} Asgekar A., Deshpande
A.A. 2001, MNRAS, 326, 1249
\bibitem[Backer(1973)]{b73} Backer D.C. 1973, ApJ, 182, 245
\bibitem[Cheng \& Ruderman(1977)]{cr77} Cheng, K., \& Ruderman, M.A. 1977, ApJ, 214, 598
\bibitem[Deshpande \& Rankin(1999)]{dr99} Deshpande, A.A., Rankin,
J.M. 1999, ApJ, 524, 1008 (DR99)
\bibitem[Deshpande \& Rankin(2001)]{dr01} Deshpande, A.A., Rankin,
J.M. 2001, MNRAS, 322, 438 (DR01)
\bibitem[Fan et al.(2001)]{fcm01} Fan G.L., Cheng K.S., \&
Manchester R.N. 2001, ApJ, 557, 297
\bibitem[Filippenko \& Radhakrishnan(1982)]{fr82} Filippenko A.V.,
Radhakrishnan V. 1982, ApJ, 263, 828
 \bibitem[Gil \& Sendyk(2000)]{gs00} Gil, J., \& Sendyk, M. 2000, ApJ, 541,
 351(GS00)
 \bibitem[Gil \& Mitra(2001)]{gm01} Gil, J., \& Mitra, D. 2001, ApJ, 550,
 383 (GM01)
\bibitem[Gil et al.(2002)]{gmm02} Gil J., Melikidze G.I., \&
 Mitra D. 2001, submitted to A\&A, astro-ph/0111474
\bibitem[Hillebrandt \& M\"{u}ller(1976)]{hm76} Hillebrandt, W., \& M\"{u}ller, E. 1976, ApJ, 207, 589
\bibitem[Karastergiou et al.(2001)]{ketal01} Karastergiou, A.,
  von Hoensbroech, A., Kramer, M., et al.
  2001, A\&A, 379, 270
\bibitem[Kijak \& Gil(1998)]{kg98} Kijak J., \& Gil J. 1998,
MNRAS, 299, 865
\bibitem[K\"ossl et al.(1988)]{kwmh88} K\"ossl D., Wolff R.G.,
M\"uller E. \& Hillebrandt W. 1988, A\&A, 205, 347
\bibitem[Melikidze, Gil \& Pataraya(2000)]{mgp00} Melikidze G.I.,
Gil J., Pataraya A.D. 2000, ApJ, 544, 1081
\bibitem[Ruderman \& Sutherland(1975)]{rs75} Ruderman, M.A., \& Sutherland, P.G. 1975, ApJ 196,
51 (RS75)
\bibitem[Sharleman, Arons \& Fawley(1978)]{saf78} Sharleman E.T., Arons J., \& Fawley W.M. 1978, ApJ, 222, 297
\bibitem[Usov \& Melrose(1995)]{um95} Usov V., Melrose D.B. 1995,
Australian J. Phys., 48, 571
\bibitem[Zhang et al.(1997)]{zql97} Zhang, B., Qiao, W., Lin, W.,
et al. 1997, ApJ, 478, 313
\bibitem[Zhang \& Harding(2000)]{zh00} Zhang B., Harding A.K.
2000, ApJ, 535, L51
\end{thebibliography}
 \end{document}